\documentclass[final,3p,times]{elsarticle}

\usepackage{amssymb,amsmath,amsfonts,mathrsfs,bm,gensymb}
\usepackage{graphicx,float,subfig}
\usepackage{lineno}

\usepackage{geometry}
\usepackage{setspace}
\usepackage{physics}
\usepackage{xcolor}
\usepackage{hyperref}

\hypersetup{
    colorlinks=true,
    linkcolor=black,
    citecolor=black,
    filecolor=black,
    urlcolor=black,
}

\makeatletter
\def\ps@pprintTitle{%
   \let\@oddhead\@empty
   \let\@evenhead\@empty
   \let\@oddfoot\@empty
   \let\@evenfoot\@oddfoot
}
\makeatother

\bibstyle{elsarticle-num}
\biboptions{sort&compress}

\begin{document}

\begin{frontmatter}

\title{Effective continuum models for the buckling of non-periodic architected sheets that display quasi-mechanism behaviors}

\author[add1]{Connor McMahan}
\author[add1]{Andrew Akerson}
\author[add2]{Paolo Celli}
\author[add3]{Basile Audoly}
\author[add1]{Chiara Daraio\corref{corr1}}\ead{daraio@caltech.edu}
\cortext[corr1]{Corresponding author}

\address[add1]{Division of Engineering and Applied Science, California Institute of Technology, Pasadena, CA 91125, USA}
\address[add2]{Department of Civil Engineering, Stony Brook University, Stony Brook, NY 11794, USA}
\address[add3]{Laboratoire de M\'{e}canique des Solides, CNRS, Institut Polytechnique de Paris, 91120 Palaiseau, France}

\begin{abstract}

\noindent In this work, we construct an effective continuum model for architected sheets that are composed of bulky tiles connected by slender elastic joints. Due to their mesostructure, these sheets feature quasi-mech\-an\-isms -- low-energy local kinematic modes that are strongly favored over other deformations. In sheets with non-uniform mesostructure, kinematic incompatibilities arise between neighboring regions, causing out-of-plane buckling. The effective continuum model is based on a geometric analysis of the sheets' unit cells and their energetically favorable modes of deformation. Its major feature is the construction of a strain energy that penalizes deviations from these preferred modes of deformation. The effect of non-periodicity is entirely described through the use of spatially varying geometric parameters in the model. Our simulations capture the out-of-plane buckling that occurs in non-periodic specimens and show good agreement with experiments. While we only consider one class of quasi-mechanisms, our modeling approach could be applied to a diverse set of shape-morphing systems that are of interest to the mechanics community.
\end{abstract}


\end{frontmatter}


\section{Introduction}
\label{s:intro}

Advanced manufacturing and synthesis technologies have given engineers the ability to design media with complex micro- and mesostructures that strongly influence bulk constitutive properties~\cite{Schaedler2011,Ware2015,Moestopo2020}. For example, the micro/mesoscale geometry can be designed to attain extreme or unconventional global mechanical behaviors such as high stiffness-to-weight ratios~\cite{Schaedler2011} and bistable auxeticity~\cite{Rafsanjani2016}. These fabrication processes have considerably expanded the design space for shape-shifting media~\cite{Klein2007, Gladman2016, Plucinsky2018} and deployable structures~\cite{Schenk2014, Boley2019}. In this context, mesoscale design has been used to create compliant features that replace conventional hinges, extensional elements and flexures~\cite{Greenberg2011, Celli2020, Ferraro2021}, or to create structures whose mechanical behaviors can be tailored by adjusting the geometry of a pattern~\cite{Guest1994,Dudte2016,Wang2017,Celli2018,Hawkes2010,Shang2018,Siefert2020, Guseinov2020,Agnelli2021}.

In structured media, the mesoscale geometry can be designed to energetically favor desired local modes of deformation~\cite{Bertoldi2017, Singh2021}. We refer to these behaviors as ``quasi-mechanisms'' when they accompany a non-negligible change in the system's energetic state. This distinguishes quasi-mechanisms from pure mechanisms, which are zero-energy kinematic modes. We emphasize that quasi-mechanisms are local behaviors: these energetic preferences can be spatially modulated by designing non-uniform mesostructures.

Within this context, origami~\cite{Greenberg2011,Dudte2016,Liu2019,Callens2017}, kirigami~\cite{Castle2014, Wang2017, Tang2017, Jiang2020} and auxetic motifs~\cite{Grima2007,Bertoldi2010,Konakovic2016,Rafsanjani2016,Konakovic2018,Celli2018} are the most popular classes of mesostructures that lead to quasi-mechanisms. 
However, demonstrations of shape-shifting materials have also been achieved using thermally responsive bilayer lattices~\cite{Guseinov2020} and in 3D structures such as snapology origami~\cite{Overvelde2017}. Quasi-mechanisms can be used to attain non-homogeneous strain field objectives (even under uniform loading conditions) by relying on non-uniform internal structures that spatially modulate local effective material properties. Morphing from a planar state to a doubly curved 3D geometry is an example of where this non-uniformity is important: Gauss' \textit{Theorema Egregium} tells us that changing a surface's Gaussian curvature requires a non-isometric mapping~\cite{Gauss1828}, which in turn requires mesostructural non-uniformity if the actuation is driven by a spatially uniform stimulus~\cite{Boley2019,Guseinov2020}. 

However, optimally designing non-uniform micro/mesostructures that lead to desired global behaviors can be challenging. The presence of geometric features at disparate length scales means that conventional finite element approaches become computationally expensive due to the need for meshes that resolve the finest features and yet span the entire structure. Homogenization theory provides a way to determine effective properties of periodic structures \cite{Allaire2012}, but in practice it is often only viable in the limited context of linear elasticity, as the presence of non-linearity and instabilities significantly complicates the methods \cite{Muller1993}. In light of this, engineers have used a variety of reduced order modeling techniques to investigate forward elastic equilibrium and stability problems, as well as to inversely design non-uniform mesostructures at a lesser computational expense. These techniques range from bar-and-hinge~\cite{Schenk2011,Filipov2017,Liu2017} and structural frame~\cite{Hayakawa2020} models that capture the mechanics of folded sheets, to representations of structural element networks that are based on effective springs~\cite{Coulais2018}, equivalent lattices~\cite{Leimer2020}, Chebyshev nets~\cite{Baek2018}, discrete elastic rods~\cite{Baek2018,Lestringant2020} and Kirchhoff rods~\cite{yu2020numerical}.

Despite the above-mentioned advancements in modeling using networks of reduced order elements, there are limitations to the existing approaches. They can be computationally expensive in cases where the structure is much larger than the mesoscale unit cell size and a reduced order element (such as a discrete elastic rod) is needed for every constituent of the physical network (e.g., in hierarchical systems). Additionally, some of these models lack the generality needed to make themselves useful to the study of other systems. For example, bar-and-hinge origami models would not be suitable for extensional spring networks. It can also be challenging to calibrate constants such that accurate results are achieved using these models. 

For these reasons, the mechanics community has pursued the development of effective continuum models. These models are powerful approaches to capturing the behavior of structures with internal geometric patterns in instances where there is a sufficient separation of length scales between the local geometric parameters and the global behaviors~\cite{Reis2018}. When this separation of scales exists, an energy density function can be constructed to capture the mechanical behaviors of the structure as if it were a bulk material, thus removing the need to resolve the geometric features at the smaller length scales with a fine mesh. This coarse meshing allows for significantly faster finite element simulations of complex physical behaviors. To this end, effective continuum models have been used to understand the behavior of periodic structured media that display quasi-mechanism behaviors~\cite{BarSinai2020} and can capture their responses to non-uniform loading conditions~\cite{Czajkowski2021,Khajehtourian2021}. However, these effective continuum modeling frameworks have not been applied to modeling the quasi-mechanism behaviors of graded media.

\begin{figure}[h]
\centering
\includegraphics[width=\textwidth]{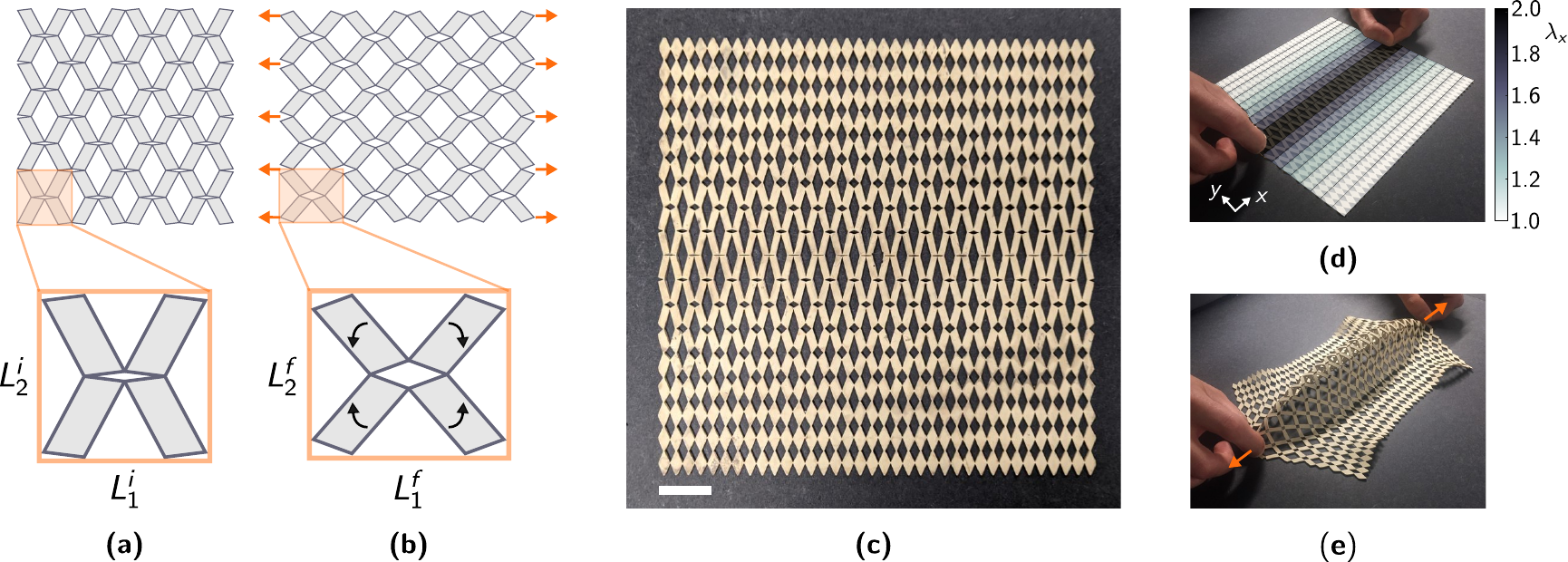}
\caption{(a-b) A sheet with a periodic cut pattern that displays a quasi-mechanism mode of deformation: rotation of tiles about slender elastic joints. As the tiles rotate, the unit cell dimensions change from $L_\alpha^i$ to $L_\alpha^f$. Although tile rotations are low-energy kinematic modes compared to other deformations, the energetic cost associated with the deformation of the joints is not negligible. (c) Introducing a gradient in the cut pattern modulates the quasi-mechanism kinematics over the sheet. The scale bar represents $3~\mathrm{cm}$. (d) The mesostructural non-uniformity shown in (c) affects the extent to which tiles can rotate in different regions of the sheet, creating kinematic incompatibilities between the quasi-mechanism behaviors of different regions. Here, $\lambda_x$ is the maximum stretch a unit cell can attain in the direction of loading through quasi-mechanism behaviors. (e) These in-plane kinematic incompatibilities lead to out-of-plane buckling. The design of the buckling sheets shown in (c-e) was first discussed in our prior work~\cite{Celli2018}.}
\label{f:intro}
\end{figure}

This article demonstrates how geometric analyses of unit cells can be used to construct effective continuum models for architected sheets with graded mesostructures. We illustrate this approach by studying generalizations of the auxetic sheets introduced by Grima \textit{et al.}~\cite{Grima2007} to spatially varying distributions of diamond-shaped cuts~\cite{Celli2018, Choi2019, Jin2020}. The tessellated unit cells consist of bulky tiles connected by slender joints, and display two elastic regimes: a soft regime that occurs when the tiles rotate about the joints (as shown in Fig.~\ref{f:intro}a-b), and a stiff regime when the joints are subjected to tension. We design heterogeneous cut patterns to provoke in-plane kinematic incompatibilities under simple point-loading scenarios, which leads to out-of-plane buckling in a region of the structure~\cite{Celli2018} (shown in Fig.~\ref{f:intro}e).

This article is organized as follows. In Section~\ref{s:model}, we discuss our effective continuum model for non-periodically patterned sheets that display quasi-mechanism behaviors. Our modeling approach entails first performing a geometric analysis of unit cells to derive their energetically favorable kinematic modes. Specifically, we derive the effect of geometric parameters on the rotational behavior of the tiles about the joints. Next, we begin constructing our strain energy density function by attributing an energy penalty to deviations from the above-mentioned kinematic modes, which may occur due to kinematic incompatibilities between neighboring regions of the sheets. Since the joints are not ideal pins, the rotation of tiles is an elastic process, albeit softer than deviations from this preferred local behavior. We use a common constitutive model for elastic materials to approximate the elastic energy associated with the tile rotations. We extract the value of a few non-geometric constants from tensile experiments on periodically patterned structures and these parameters are then used to simulate the non-periodic structure. This type of effective material modeling enables us to use a coarse mesh to solve for pre-buckled equilibrium, the onset of instabilities, and post-buckled equilibrium. The numerical approach is discussed in Section~\ref{s:numerics}, and we compare these numerical results to a new set of experiments in Section~\ref{s:results}, highlighting the good agreement between coarse mesh finite element simulations and experiments. Our concluding remarks and perspective for future work are presented in Section~\ref{s:conclusions}. While our modeling method is demonstrated for the class of quasi-mechanisms discussed above, we believe it would be straightforward to apply it to many other quasi-mechanisms that are of interest to the mechanics community, such as origami tessellations~\cite{Callens2017} and shape-shifting bilayer lattices~\cite{Guseinov2020}.


\section{Modeling approach}
\label{s:model}

In this section, we discuss how a strain energy density function can be extracted by modeling the effect that mesoscale geometric features have on a structure's energetically favorable local modes of deformation. Our approach is presented for modeling effective continua within the context of initially flat sheets with diamond-shaped cut patterns, although it could be generalized to other types of 2D or 3D architected media.

\subsection{Quasi-mechanism kinematics}\label{s:QMkin}

\begin{figure}[!htb]
\centering
\includegraphics[scale=1.1]{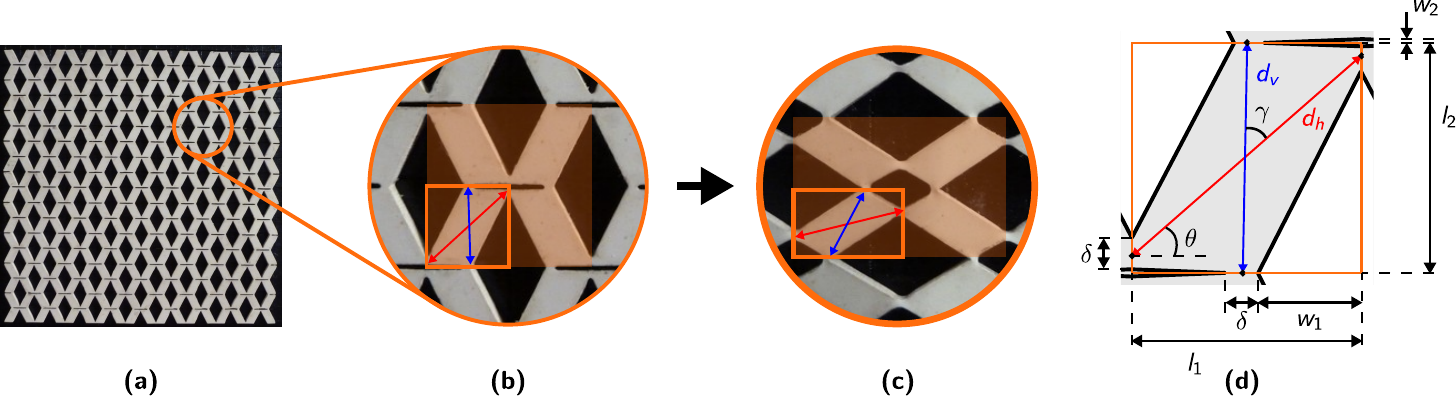}
\caption{(a) An example of a sheet with a uniform pattern of diamond-shaped cuts. (b) A unit cell (shaded) consists of four tiles (boxed). (c) The quasi-mechanism kinematics consist of tile rotations about the slender elastic joints. This deformation mode can be entirely described by the projection of the tile diagonals onto the fixed orthogonal coordinate frame. This rotational mode has a non-negligible energetic cost, but one that is still much lesser than deformations where the joints are under tension or shear. (d) The reference configuration of the boxed tile shown in (b). Five parameters define the geometry of a unit cell: $l_1$ and $l_2$ are the reference configuration lengths of the unit cell grid spacing in the $\mathbf{e}_1$ and $\mathbf{e}_2$ directions, $\delta$ is the width of the slender joints, and $w_1$ and $w_2$ are the half-widths of the two diamond-shaped cuts that define the tiles' inclinations. The diagonals $d_v$ and $d_h$ and the angle $\gamma$ between these two can be computed from those parameters. Finally, $\theta$ is the angle between the red diagonal, $d_h$, and the $\mathbf{e}_1$ direction. As the tile rotates from one configuration to another, this angle varies (as shown in b-c). The projected lengths of the tile's deformed configuration in the $\mathbf{e}_1$ and $\mathbf{e}_2$ directions are $d_h\cos(\theta)$ and $d_v\sin(\gamma+\theta)$, respectively. This allows us to compute the unit cell stretches: only the rotation of one tile about a joint needs to be analyzed to determine the quasi-mechanism kinematics of the unit cell. (a-d) Adapted from~\cite{Celli2018} by permission of The Royal Society of Chemistry.}
\label{f:kin}
\end{figure}

Our aim is to create an effective continuum model that captures the quasi-mechanism kinematics of sheets with diamond-shaped cut patterns (Fig.~\ref{f:kin}a). These sheets are tessellations of unit cells that are composed of four bulky tiles connected by slender elastic joints (Fig.~\ref{f:kin}b). The structures may be either periodic or non-periodic tessellations of unit cells (as in Fig.~\ref{f:kin}a or Fig.~\ref{f:intro}c, respectively). In either case, the quasi-mechanism local modes of deformation can be derived from a simple geometric analysis relating unit cell geometry to the rigid body rotations of the bulky tiles about the joints (Fig.~\ref{f:kin}b-c).

Five spatially varying geometric parameters constitute a geometry vector field $\boldsymbol{\phi}(x_\alpha)$ and define the quasi-mech\-an\-ism kinematics of our sheets. Namely $\boldsymbol{\phi}=\{l_1,l_2,\delta,w_1,w_2\}$, where $l_1(x_\alpha)$ and $l_2(x_\alpha)$ are the lengths of the unit cell grid spacing in the $\mathbf{e}_1$ and $\mathbf{e}_2$ directions, $\delta(x_\alpha)$ is the width of the slender joints, and $w_1(x_\alpha)$ and $w_2(x_\alpha)$ are the half-widths of the two diamond-shaped cuts that define the tiles' inclinations. These parameters are illustrated in Fig.~\ref{f:kin}d. A few geometric parameters that are functions of the five mentioned above are also shown in Fig.~\ref{f:kin}d and will be discussed below. 

We seek to identify a function $g(\mathbf{C}, \boldsymbol{\phi})$ such that the local quasi-mechanisms are described by the implicit relation $g(\mathbf{C}, \boldsymbol{\phi})=0$. Here, $\mathbf{C}$ is the right Cauchy-Green strain tensor. 
To do so, we first define a unit cell as a $2 \times 2$ arrangement of quadrilateral tiles. Due to the symmetry of the unit cell, we can fully describe its quasi-mechanism kinematics by analyzing the geometry and rotation of a single tile. We use the bottom left tile in the unit cell, such as the one boxed in Fig.~\ref{f:kin}b-c. For a unit cell located at $x_\alpha$ with geometry defined by $\boldsymbol{\phi}(x_\alpha)=\{l_1(x_\alpha),l_2(x_\alpha),\delta(x_\alpha),w_1(x_\alpha),w_2(x_\alpha)\}$, the respective lengths $d_h$ and $d_v$ of the diagonals illustrated in Fig.~\ref{f:kin}d in red and blue are

\begin{equation} d_h(\boldsymbol{\phi})=\sqrt{l_1^2+(l_2-2w_2-\delta)^2}\,\,\,\,\,\,\mathrm{and}\,\,\,\,\,\,d_v(\boldsymbol{\phi})=\sqrt{l_2^2+(l_1-2w_1-\delta)^2}\,.
\end{equation}

\noindent The angle $\gamma$ between these two diagonals is given in terms of the geometric parameters $\boldsymbol{\phi}$ as

\begin{equation} 
\gamma(\boldsymbol{\phi}) = \frac{\pi}{2}-\arctan\left(\frac{l_2-2w_2-\delta}{l_1}\right)-\arctan\left(\frac{2w_1+\delta-l_1}{l_2}\right)\,.
\end{equation}

As the tile rotates about the joint, the angle $\theta$ between the diagonal $d_h$ and the $\mathbf{e}_1$ direction varies, as shown in Fig.~\ref{f:kin}b-c. During this tile rotation, the projected lengths of the tile diagonals on the fixed orthogonal frame $\mathbf{e}_i$ change, and the unit cell will have effective stretches $\lambda_1$ and $\lambda_2$ of 
\begin{equation}\label{eq:lamTh}
\lambda_1(\theta)=\frac{d_h\cos\theta}{l_1}\,\,\,\,\,\,\mathrm{and}\,\,\,\,\,\,\lambda_2(\theta)=\frac{d_v\sin(\gamma+\theta)}{l_2}\,.
\end{equation}
We can invert the function for $\lambda_1(\theta)$ to obtain  $\theta(\lambda_1)$ as
\begin{equation}\label{eq:thlamx}
\theta(\lambda_1)=\arccos\left( \frac{\lambda_1 l_1}{d_h} \right)\,.
\end{equation}
Substituting \eqref{eq:thlamx} into the expression for $\lambda_2(\theta)$ in \eqref{eq:lamTh} leads to the following explicit formula for $\lambda_2(\lambda_1)$:
\begin{equation}\label{eq:explicitKin}
\lambda_2(\lambda_1)=\frac{d_v}{l_2}\sin\left[\gamma+\arccos\left( \frac{\lambda_1 l_1}{d_h} \right)\right]\,.
\end{equation}

\noindent We first derived this explicit function for the quasi-mechanism kinematics in our prior work~\cite{Celli2018}. Through trigonometric identities and algebraic manipulation, this can be written in implicit form:

\begin{equation}\label{eq:implicitKin}
    \Bigg(\frac{l_1\lambda_1}{d_h}\Bigg)^2+\Bigg(\frac{l_2\lambda_2}{d_v}\Bigg)^2-2\sin(\gamma)\frac{l_1 \lambda_1}{d_h}\frac{l_2 \lambda_2}{d_v}-\cos^2(\gamma)=0
\end{equation}

\noindent In our reference frame, the implicit function \eqref{eq:implicitKin} can be rewritten using the components of $\mathbf{C}$, since $\mathbf{C_{11}}=\lambda_1^2 \mathbf{e}_1\otimes\mathbf{e}_1$ and $\mathbf{C_{22}}=\lambda_2^2\mathbf{e}_2\otimes\mathbf{e}_2$:

\begin{equation}\label{eq:mechC}
    g(\mathbf{C}, \boldsymbol{\phi})=\frac{l_1^2\mathbf{C_{11}}}{d_h^2(\boldsymbol{\phi})}+\frac{l_2^2\mathbf{C_{22}}}{d_v^2(\boldsymbol{\phi})}-2\sin\Big(\gamma(\boldsymbol{\phi})\Big)\frac{l_1 l_2}{d_h(\boldsymbol{\phi}) d_v(\boldsymbol{\phi})}\sqrt{\mathbf{\det C}}-\cos^2\Big(\gamma(\boldsymbol{\phi})\Big)=0
\end{equation}

The quasi-mechanism kinematics expressed in \eqref{eq:mechC} describe the unit cells' preferred modes of local deformation as a function of geometric parameters. We emphasize that a unit cell may not deform according to this function. For example, this may occur if neighboring unit cells have a different geometry and cause kinematic incompatibility or if global loading conditions make these modes of deformation energetically unfavorable. In these cases, $g(\mathbf{C},\boldsymbol{\phi})\neq0$. In Section~\ref{s:StrEnFun}, we will model the stiffening that occurs when \eqref{eq:mechC} cannot be satisfied by embedding this kinematic description as a penalty term in our strain energy function. 

\subsection{Kinematics of a thin elastic plate}
\label{s:conf}

Our aim is to embed the quasi-mechanism behavior described by \eqref{eq:mechC} into an effective continuum model. We consider a thin elastic plate whose material particle positions of the mid-plane in an initially flat reference configuration are $\mathbf{X}=x_\alpha \mathbf{e}_\alpha$. The indices $\alpha$ and $\beta$ in this subsection relate to the mid-plane of the plate (we use the Einstein summation convention for repeated indices), and the index `3' corresponds to the direction normal to the reference surface. The coordinate frame $\{\mathbf{e}_i\}$ is fixed and orthonormal. The domains for the material coordinates $x_\alpha$ are $x_1\in[0,a]$ and $x_2\in[0,b],$ where $a$ and $b$ are constants. The thickness $t$ is much smaller than the other material domain dimensions, and we seek the mid-surface mapping $\mathbf{\boldsymbol{\chi}}(x_\alpha)$:

\begin{equation}\label{eq:mapping}
\boldsymbol{\chi}(x_\alpha)=\Big(x_\alpha+u_\alpha(x_\beta)\Big)\mathbf{e}_\alpha+w(x_\beta)\mathbf{e}_3,
\end{equation}

\noindent where $u_\alpha$ and $w$ are the in-plane and out-of-plane components of the mid-plane displacement vector, respectively. The deformation gradient tensor $\mathbf{\tilde{F}}=\nabla \boldsymbol{\chi}$ can be expressed in terms of the gradients of $u_\alpha$ and $w$. We label $\mathbf{F}$ as the in-plane component of the deformation gradient tensor ($\mathbf{F}\equiv\mathbf{I}+\nabla u_\alpha)$. Since we have two material coordinates embedded in three spatial dimensions, the deformation gradient assumes the following form:

\begin{equation}\label{eq:defGrad}
\mathbf{\tilde{F}}=
\begin{bmatrix}
1+u_{1,1} & u_{1,2} \\
u_{2,1} & 1+u_{2,2} \\
w_{,1} & w_{,2} \\
\end{bmatrix} = 
\begin{bmatrix}
\mathbf{F}\\
\nabla w\\
\end{bmatrix}
\end{equation}

We use the right Cauchy-Green deformation tensor, $\mathbf{C}$, as our measure for in-plane strain, and the Laplacian of the out-of-plane deflections, $\Delta w$, as our bending strain measure: 

\begin{equation}\label{eq:rightCauchy}
\mathbf{C}=\mathbf{\tilde{F}^T \tilde{F}}=\mathbf{F^T F} + \nabla w \otimes \nabla w,\ \ \ \ \ \ \Delta w = \frac{\partial^2 w}{\partial x_1^2}+\frac{\partial^2 w}{\partial x_2^2}
\end{equation}

\subsection{Strain energy}\label{s:StrEnFun}

Now that we have an implicit function \eqref{eq:mechC} describing the quasi-mechanism behavior and a formulation of thin plate kinematics, we can construct a strain energy density function for our sheets. The first step is to attribute an energy penalty $\Psi_p$ for deviations from the quasi-mechanism behavior. As discussed in Section~\ref{s:QMkin}, $g(\mathbf{C},\boldsymbol{\phi})=0$ when local deformations correspond to quasi-mechanism behaviors, and $g(\mathbf{C},\boldsymbol{\phi})\neq0$ when there is a deviation from these energetic preferences. Therefore we can write our energy penalty $\Psi_p$ as

\begin{equation}\label{eq:energyPen}
    \Psi_p=\frac{1}{2\eta}g^2(\mathbf{C}, \boldsymbol{\phi})\ ,
\end{equation}

\noindent where $\eta$ is a small parameter. For our perforated sheets, $g(\mathbf{C},\boldsymbol{\phi})$ is given in \eqref{eq:mechC}. Therefore,

\begin{equation}\label{eq:energyPenQM}
\Psi_p=\frac{1}{2\eta}\Bigg( \frac{l_1^2\mathbf{C_{11}}}{d_h^2}+\frac{l_2^2\mathbf{C_{22}}}{d_v^2}-2\sin(\gamma)\frac{l_1 l_2}{d_h d_v}\sqrt{\mathbf{\det C}}-\cos^2(\gamma) \Bigg)^2\ .
\end{equation}

For elastic bodies, deforming according to these preferential modes will still entail non-zero energy. Thus, we must also assign a soft elastic energy density $\Psi_s$ to this scenario (this softness is relative to the energy expense of deviating from quasi-mechanism behaviors). A compressible Neo-Hookean model provides the flexibility to approximate our experimental data from tensile tests well while using only two material parameters. Therefore, the total membrane strain energy density function $\Psi_m(\mathbf{C}, \boldsymbol{\phi}) = \Psi_p(\mathbf{C}, \boldsymbol{\phi}) + \Psi_s(\mathbf{C})$ is

\begin{equation}\label{eq:memEnergy}
    \Psi_m(\mathbf{C}, \boldsymbol{\phi}) = \Psi_p(\mathbf{C}, \boldsymbol{\phi})+\frac{\mu}{2}(\bar{I}_1-2) + \frac{\lambda}{2}(J-1)^2\ ,
\end{equation}

\noindent where $J=\sqrt{\det(\mathbf{C})}$, $\bar{I}_1=\mathrm{tr} (\mathbf{C}) J^{-1}$, $\mu$ and $\lambda$ are the Lam\'{e} parameters and $\Psi_p$ is given in \eqref{eq:energyPenQM}. Our bending energy density functions is

\begin{equation}\label{eq:bendEnergy}
\Psi_b= \frac{B (\Delta w)^2}{2}\ ,
\end{equation}

\noindent where $B$ is a bending stiffness constant. Our strain energy per unit thickness is the sum of $\Psi_m$ and $\Psi_b$, integrated over the 2D domain spanned by the mid-plane of the sheet, $\Omega$:

\begin{equation} \label{eq:energy}
    \mathcal{E}(\mathbf{u}, w) =\int_{\Omega} \bigg(\Psi_m(\mathbf{C}, \boldsymbol{\phi}) + \Psi_b(\Delta w)\bigg) \ dA
\end{equation}

All of the parameters in the energy function are either geometric or can be extracted from three simple tensile experiments: one on a dogbone specimen of the bulk rubber with no cut patterns, and two (conducted in orthogonal directions) on a sheet with periodic but anisotropic cuts.

\subsection{Contact model}

In the case where the sheet lies on a rigid surface, we wish to enforce the contact condition $w \geq 0$. While techniques such as the active set method directly impose this constraint, we opt to relax this condition and instead use a rather simple penalty-based contact model. Thus, for problems where the sheet is lying on a flat surface, we consider a contact penalty energy for negative out-of-plane deflections:
\begin{equation}
    \Psi_c(w) = \frac{P}{2}(dw^-)^2, \qquad dw^- = \min(0, w+\varepsilon),
\end{equation}
where $P$ is the penalty stiffness and $\varepsilon > 0$ is a small tolerance length. Notice that the contact energy is nonzero only when $w < -\varepsilon$. This ensures that the contact condition does not interfere with the stability of the initially flat, unbuckled plate, and only becomes active post-bifurcation. We add this contact energy onto \eqref{eq:energy} to give the total energy functional
\begin{equation}\label{eq:energyWpen}
    \mathcal{E}(\mathbf{u}, w) = \int_\Omega \Psi_m(\mathbf{C},\boldsymbol{\phi}) + \Psi_b(\Delta w) + \Psi_c(w) \ dA.
\end{equation}

\noindent We will discuss the variations of this energy to compute equilibrium and stability in Section~\ref{s:numerics}.

\section{Finite element implementation}\label{s:numerics}

In this section, we present the equilibrium conditions for the system. Using a mixed formulation, we compute the solution using standard first order Lagrange polynomial finite elements. More details for our solution procedure and stability analysis are provided in the appendices. We implement this formulation in the deal.II open source finite element library~\cite{dealii}. \begin{figure}[!htb]
\centering
\includegraphics[width=0.6\textwidth]{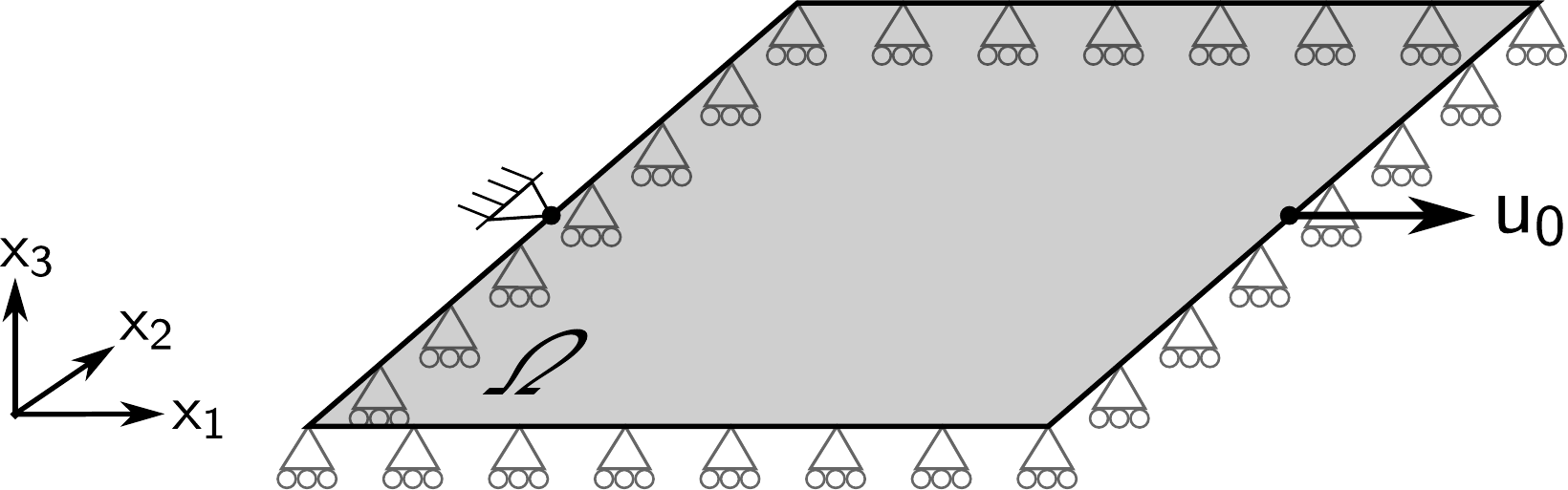}
\caption{An example of a domain and of a set of boundary conditions used in our simulations. In-plane displacements are prescribed on a portion of the boundary and in-plane traction-free edges are observed on the remainder. Additionally, we constrain out-of-plane displacements and have no applied moments on the entire boundary. This drawing displays the boundary conditions used to model the sheet with non-uniform cut patterns shown in Fig~\ref{f:intro}c-e.}
\label{f:comp_domain}
\end{figure}

We consider a rectangular domain in a displacement-controlled setting. The in-plane displacements $\mathbf{u}$ are prescribed on $\partial_u \Omega \subset \partial \Omega$ and we have in-plane traction free edges on the remainder, $\partial_f \Omega = \partial \Omega \backslash \partial_u \Omega$. Additionally, we constrain out-of-plane displacements $w$ and have moment-free edges on the entire boundary. Fig~\ref{f:comp_domain} shows an example of a domain and of a set of boundary conditions used in some of our simulations. While the boundary conditions may be altered for a more general case, the mixed formulation discussed in Subsection~\ref{s:MixedForm} may not be appropriate for situations such as clamped boundaries. 

\subsection{Equilibrium and mixed formulation}\label{s:MixedForm}

The equilibrium condition is the stationarity of our energy functional from \eqref{eq:energyWpen} in both $\mathbf{u}$ and $w$, 
\begin{equation}\label{eq:stat}
	\frac{\mathrm{d}}{\mathrm{d} \kappa} \Big[ \mathcal{E}(\mathbf{u} + \kappa \delta \mathbf{u}, w + \kappa \delta w) \Big]_{\kappa = 0} = 0 \qquad \text{for all} \quad  \delta \mathbf{u} \in \mathcal{U}_0,\quad  \delta w \in H^2_0(\Omega),
\end{equation}
where $\mathcal{U}_0$ is the set of kinematically admissible in-plane displacement variations
\begin{equation}
	\mathcal{U}_0 = \left \{ \mathbf{u} \in \left( H^1(\Omega) \right)^2, \ \mathbf{u} = 0 \ \text{on} \ \partial_u \Omega \right \},
\end{equation}
and we search for solutions $\mathbf{u} \in \mathcal{U}$ and $w \in \mathcal{W}$ where
\begin{equation}
    \mathcal{U} = \left \{ \mathbf{u} \in \left( H^1(\Omega) \right)^2, \ \mathbf{u} = \mathbf{u}_0 \ \text{on} \ \partial_u \Omega \right \}, \quad	\mathcal{W} = \left \{ w \in H^2(\Omega), \ w = w_0 \ \text{on} \ \partial \Omega \right \}.
\end{equation}
A common issue for plate problems is the bi-harmonic operator on $w$ that arises from the Gateaux derivative of the bending energy. In this case, the weak form contains a product of the second derivative of $w$ and its variation, so that the usual Galerkin finite element method with even quadratic Lagrange polynomial shape functions is not appropriate.\footnote{Standard Lagrange polynomial shape functions have discontinuous first-derivatives at the boundaries of elements. This would result in integrating the product of two Dirac delta functions, which is undefined.} Therefore, we turn to a mixed formulation that is widely used for linear biharmonic problems~\cite{Boffi2013}. We introduce a scalar function $v \in H^1_0(\Omega)$ and set it equal to $\Delta w$ by considering an augmented energy
\begin{equation}
	\widehat{\mathcal{E}}(\mathbf{u}, w) =  \sup_{v \in H^1_0(\Omega)} \ \int_{\Omega}  \Psi_m(\mathbf{C}) + \Psi_c(w)  - B \left( \nabla w \cdot \nabla v - \frac{1}{2} \abs{v}^2 \right) \ dA.
\end{equation}
Stationarity of $\widehat{\mathcal{E}}$ in both $\mathbf{u}$ and $w$, along with the suprema condition on $v$, gives the weak form of equilibrium
\begin{equation} \label{eq:mixed_eq_weak}
	\begin{aligned}
		0 &= \int_{\Omega} \left( 2 \mathbf{F} \pdv{\Psi_m}{\mathbf{C}} \right) : \nabla \delta \mathbf{u} \ dA  \qquad && \forall \delta  \mathbf{u} \in \mathcal{U}_0, \\
		0 &= \int_{\Omega} \left( 2 \pdv{\Psi_m}{\mathbf{C}}\nabla w \right) \cdot \nabla \delta w + \pdv{\Psi_c}{w}  - B \nabla v \cdot \nabla \delta w  \ dA \qquad && \forall \delta w \in H^1_0 (\Omega), \\
		0 &= \int_{\Omega} - B \nabla w \cdot \nabla \delta v  - B v \delta v   \ dA \qquad && \forall \delta v \in H^1_0(\Omega). \\
	\end{aligned}
\end{equation}
The first two lines in \eqref{eq:mixed_eq_weak} are the equilibrium relations for in-plane and out-of-plane displacements, respectively. The final line is the constraint that $v = \Delta w$ weakly. The strong form of these relations can be found in~\ref{app:strong}. Notice that \eqref{eq:mixed_eq_weak} only contains first derivatives of the displacements and their variations. It is shown in \cite{Boffi2013} that we may now consider $w \in H^1(\Omega)$. Therefore, we use a Galerkin finite element formulation with p = 1 shape functions for the fields $\mathbf{u}$, $w$ and $v$. We solve the nonlinear system with typical Newton-Raphson iterations. Details on the finite element formulation and solution procedure can be found in~\ref{app:fem}. 

\subsection{Stability analysis}
To probe the stability of an equilibrium configuration, it is common practice to calculate the eigenvalues of the tangent stiffness matrix. A negative eigenvalue implies an instability, and the equilibrium solution can then be perturbed in the direction of the corresponding eigenvector to explore the buckled solution. However, the mixed formulation complicates this procedure. To assess stability, we must restrict the eigenvectors to the subspace upon which the constraint $v = \Delta w$ is satisfied. To this end, we consider an effective stiffness matrix on this subspace. By solving the linear constraint explicitly, we can condense $v$ out of the system matrix. Then, we calculate eigenvalues of this reduced stiffness matrix to asses stability. We use the linear constraint to map the corresponding eigenvector back to the full variable set and perturb the system. The magnitude of the perturbation is chosen to be on the same order as the displacement increment. The direction of the perturbation is decided such that the $w$ component at the middle of the sheet is positive. The full details of the stability analysis can be found in~\ref{app:stab}. 

\section{Results}
\label{s:results}

In this section, we discuss the extraction of effective material model constants from experiments on sheets with uniform cut patterns and we compare experimental and numerical results on the post-buckling behavior of sheets with non-periodic mesostructure. 

\subsection{Extracting model constants from experiments on sheets with uniform cut patterns}\label{s:constantExtraction}

As discussed in Section~\ref{s:StrEnFun}, our energy given in \eqref{eq:energyWpen} requires the extraction of four parameters from experiments: the Lam\'{e} moduli ($\lambda$ and $\mu$), the energy penalty parameter ($\eta$), and the bending stiffness ($B$). We obtained $\lambda$, $\mu$ and $\eta$ from tensile tests on the specimen with uniform cut patterns shown in Fig.~\ref{f:kin}a, where $l_1=l_2=6~\mathrm{mm}$, $\delta=l_1/8$, $w_1=(l_1-\delta)/2$, and $w_2=0~\mathrm{mm}$. The sheets have a thickness of $t=1.55~\mathrm{mm}$, width dimensions of $108~\mathrm{mm}$ in each direction and are made of natural rubber gum. The diamond-shape cuts were made using a laser cutter.

\begin{figure}[!htb]
\centering
\includegraphics[width=0.4245\textwidth]{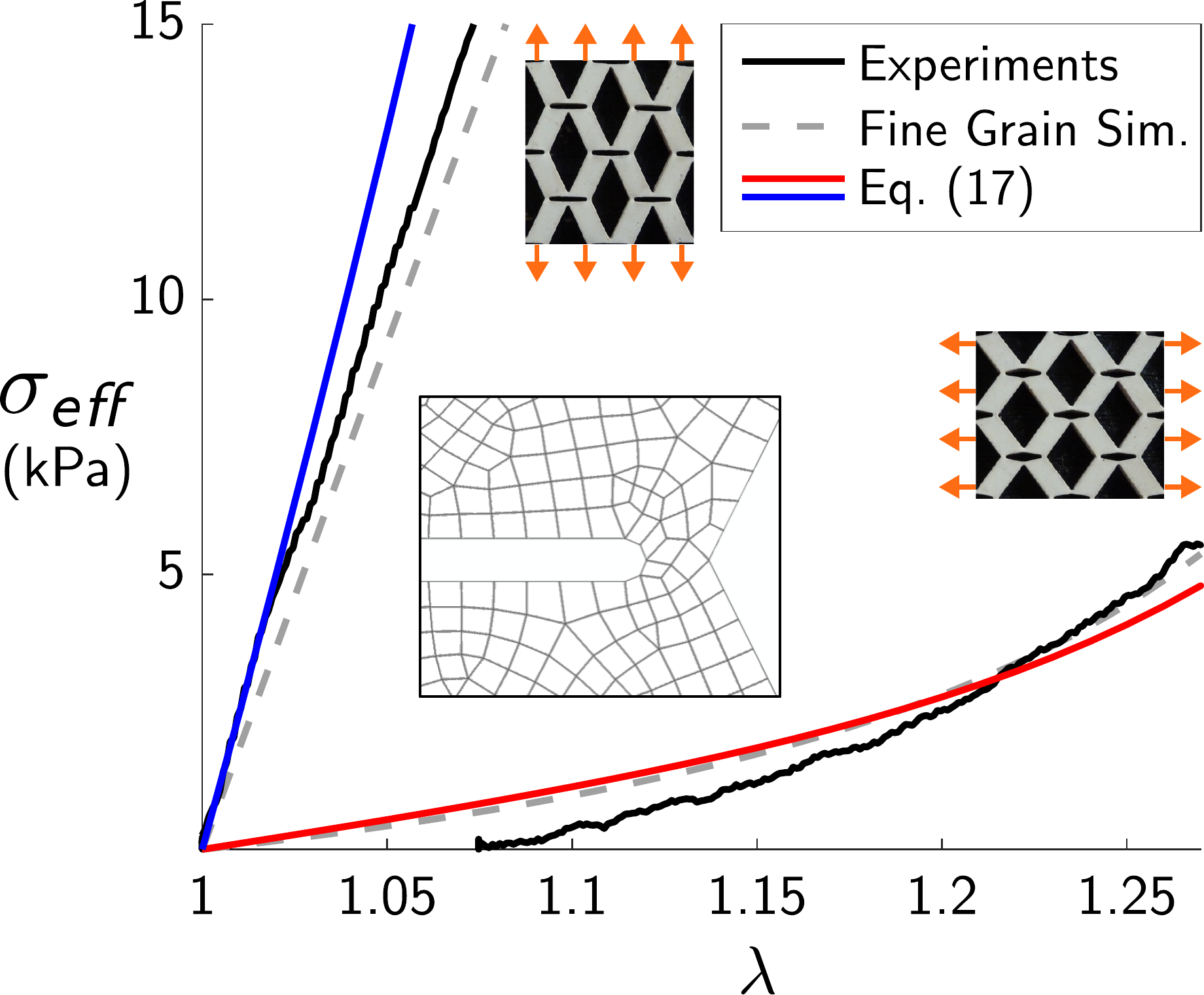}
\caption{Effective stress vs.~stretch for a sheet with a periodic cut pattern. The insets show four unit cells of this structure, see Fig.~\ref{f:kin}a for an image of the entire sheet. We compare our effective continuum model (solid red and blue lines) represented by \eqref{eq:energyWpen} to experiments (solid black lines) and fine-grain finite element simulations (gray dashes) that fully resolve the small geometric features in our sheets. These experiments and the fine-grain simulations (using Abaqus/Standard) were conducted in our prior work~\cite{Celli2018}). The experimental curve for the soft loading direction does not start at $\lambda=1$ due to the effect of gravity in a vertically loaded tensile testing machine. The inset on the bottom left of the figure shows a small region of the mesh used in the Abaqus simulations to capture the geometry of the elastic joints. The large number of elements needed for these fine grain simulations motivates the usage of effective continuum models. The insets in this image were adapted from~\cite{Celli2018} by permission of The Royal Society of Chemistry.}
\label{f:StrStr}
\end{figure}

The specimen was placed on a custom apparatus that grips the edges with roller pins, thus allowing free sliding in the direction perpendicular to the tension. To obtain $\lambda$ and $\mu$, the sheet was loaded in the direction that induces quasi-mechanism behavior (rotation of the tiles about the elastic joints). Since the sheet's cut pattern is uniform, no kinematic incompatibilities arise and only the soft elastic mode is present. The values of $\mu=17~\mathrm{kPa}$ and $\lambda = 0.1~\mathrm{kPa}$ provided a good fit to our data, as shown in Fig.~\ref{f:StrStr}. To attain $\eta$, the sheet was loaded in the perpendicular direction, where tiles do not rotate because their diagonals are aligned in the direction of loading and the elastic joints are in tension. We attain a good fit of our data by setting $\eta=0.002~\mathrm{kPa}^{-1}$. Fig.~\ref{f:StrStr} shows a comparison of effective continuum simulations of the in-plane elastic behaviors with experiments and Abaqus/Standard simulations from prior work~\cite{Celli2018}, where the mesh fully resolves the fine features of the specimen geometry.

We adjust the classic bending stiffness for a Kirchhoff-Love plate~\cite{Timoshenko1959} by including a scaling factor $\alpha(f)$ that accounts for the reduced bending stiffness of a sheet with porosity $f$. Therefore, the bending stiffness of the patterned sheet can be written in the following form:

\begin{equation}
    B = \frac{\alpha(f) E t^2}{12(1-\nu^2)}\ .
\end{equation}
\noindent Here, $E=2~\mathrm{MPa}$ is Young's modulus (obtained from linear regime tensile tests on a $55~\mathrm{mm}\times9.2~\mathrm{mm}\times1.5~\mathrm{mm}$ dogbone sample of natural rubber), $t$ is the sheet thickness, and $\nu=0.5$ is Poisson's ratio. A recent paper by Shrimali, et al.~\cite{Shrimali2021} showed that the effective bending stiffness of thin perforated plates is much more dependent on the plate's porosity $f$ than on the shape or size of the perforations. This holds both for plates where the sheet thickness is much smaller than the unit cell dimension, and vice versa. Given the porosity of our sheets ($f\approx0.5$), we adopt a scaling value of $\alpha(f)=0.25$, as suggested by the results in~\cite{Shrimali2021}. Their results also justify our use of a uniform bending stiffness. Again, \eqref{eq:energyWpen} is the strain energy per unit thickness, hence the scaling of $B$ with $t^2$. Based on these considerations, no additional experiment is required to obtain the bending stiffness.

\subsection{Out-of-plane buckling of sheets with graded mesostructure}
We now consider a more interesting pattern of cuts that is non-periodic, and where spatial variations in the local quasi-mechanism behavior lead to kinematic incompatibilities. To model the behavior of these sheets, we update the geometry vector $\boldsymbol{\phi}(x_\alpha)=\{l_1(x_\alpha),l_2(x_\alpha),\delta(x_\alpha),w_1(x_\alpha),w_2(x_\alpha)\}$. We have three specimens of equal thickness $t=1.55$~mm, but varying aspect ratios. Now, $l_1=\{ 4.5~\mathrm{mm},\ 6~\mathrm{mm},\ 7.5~\mathrm{mm}\}$ for the three sheets (the overall width dimensions of the square sheets scale linearly with $l_1$ to $162~\mathrm{mm}$, $216~\mathrm{mm}$, and $270~\mathrm{mm}$, respectively). The other parameters are $l_2=2l_1$, $\delta = l_1/8$, $w_1=(l_1-\delta)/2$, and $w_2(x_\alpha)=\frac{l_1-\delta}{2}\Big(1-\sin\frac{\pi x_2}{18l_2}\Big)$. The non-uniform geometry is accounted for by considering spatially varying $w_2(x_\alpha)$ in the finite element formulation. We note that although the geometric parameter $w_2(x_\alpha)$ is non-uniform, we still use a uniform soft elastic energy density, $\Psi_s$, because it represents the energetic cost of the non-ideal mechanism and the joint density is still uniform. 

\begin{figure}[h]
\centering
\includegraphics[width=\textwidth]{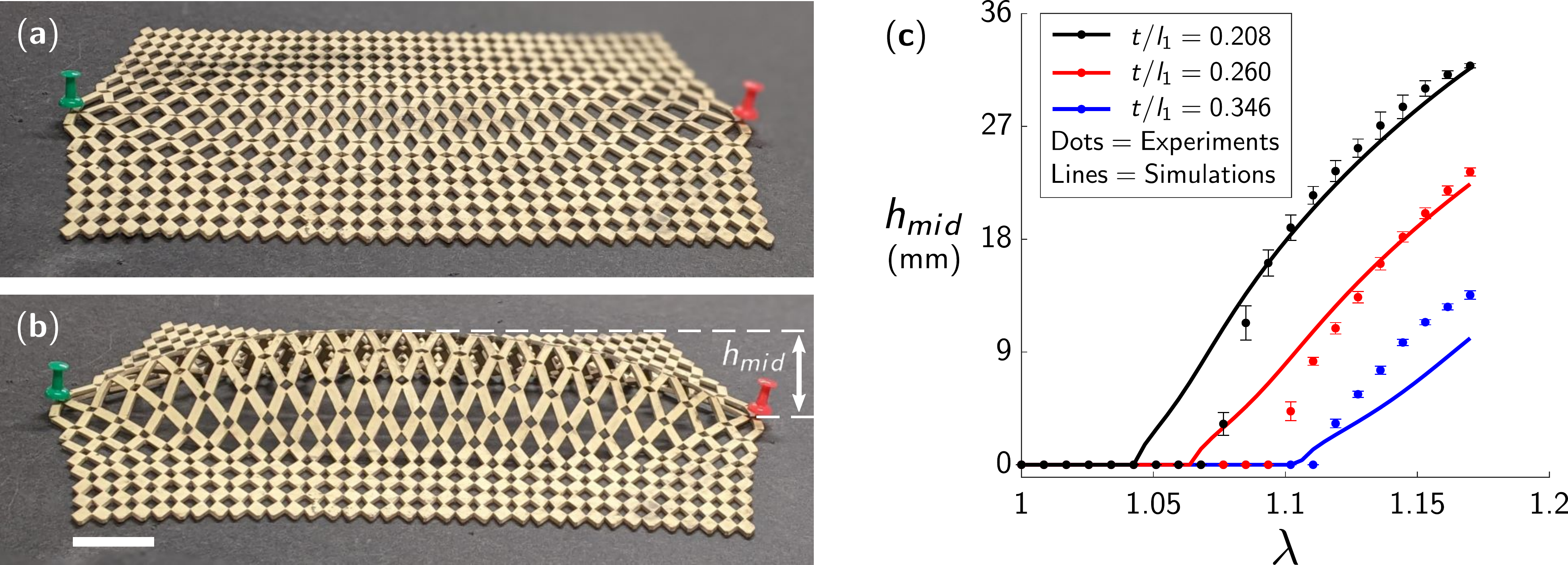}
\caption{Buckling behavior of sheets with non-uniform cut patterns. (a) Up to a certain stretch $\lambda$, point displacements lead to in-plane deformations. (b) Following a critical value of $\lambda$, the in-plane kinematic incompatibilities will lead to out-of-plane buckling. The scale bar represents 3~cm. (c) Comparison of dome height between effective continuum simulations (solid lines) and experiments (dots) for sheets of three aspect ratios. Here, $h_{mid}$ is the height of a sheet's center point, $\lambda$ is the stretch of the sheet's center line in the $\mathbf{e}_1$ direction, $t$ is the sheet thickness, and $l_1$ is the length of the unit cell grid spacing in the $\mathbf{e}_1$ direction.}
\label{f:heights}
\end{figure}

The geometric gradation of the mesostructure leads to variations in the local quasi-mechanism behavior over the extent of the sheet. This causes in-plane kinematic incompatibilities, which lead to out-of plane buckling after each sheet's critical stretch is reached, as shown in Fig.~\ref{f:heights}a-b. We show the buckled mode nucleation and the evolution of the post-buckled height of the central point in the sheets as a function of boundary point displacement in Fig.~\ref{f:heights}c. We compare simulations of our effective continuum model (computed using the deal.II finite element library~\cite{dealii} on a $36 \times 36$ uniform quadrilateral mesh) to measurements of the physical samples (using a level-calibrated mounted caliper) and see excellent agreement between the two, especially at larger boundary displacements. As expected, the stretch at which buckling occurs is delayed by increasing the thickness-to-width ratio. The difference between the computational predictions and experimental measurements of buckling nucleation and height at lower stretch values can be partially attributed to the fact that our simulations do not account for friction with the table or gravity. These two physical processes are important since the material is soft and bending is a low-energy deformation for shells with small gaussian curvature. As the dome height increases, the structure becomes less susceptible to the effect of gravity.

\begin{figure}[!h]
\centering
\includegraphics[width=\textwidth]{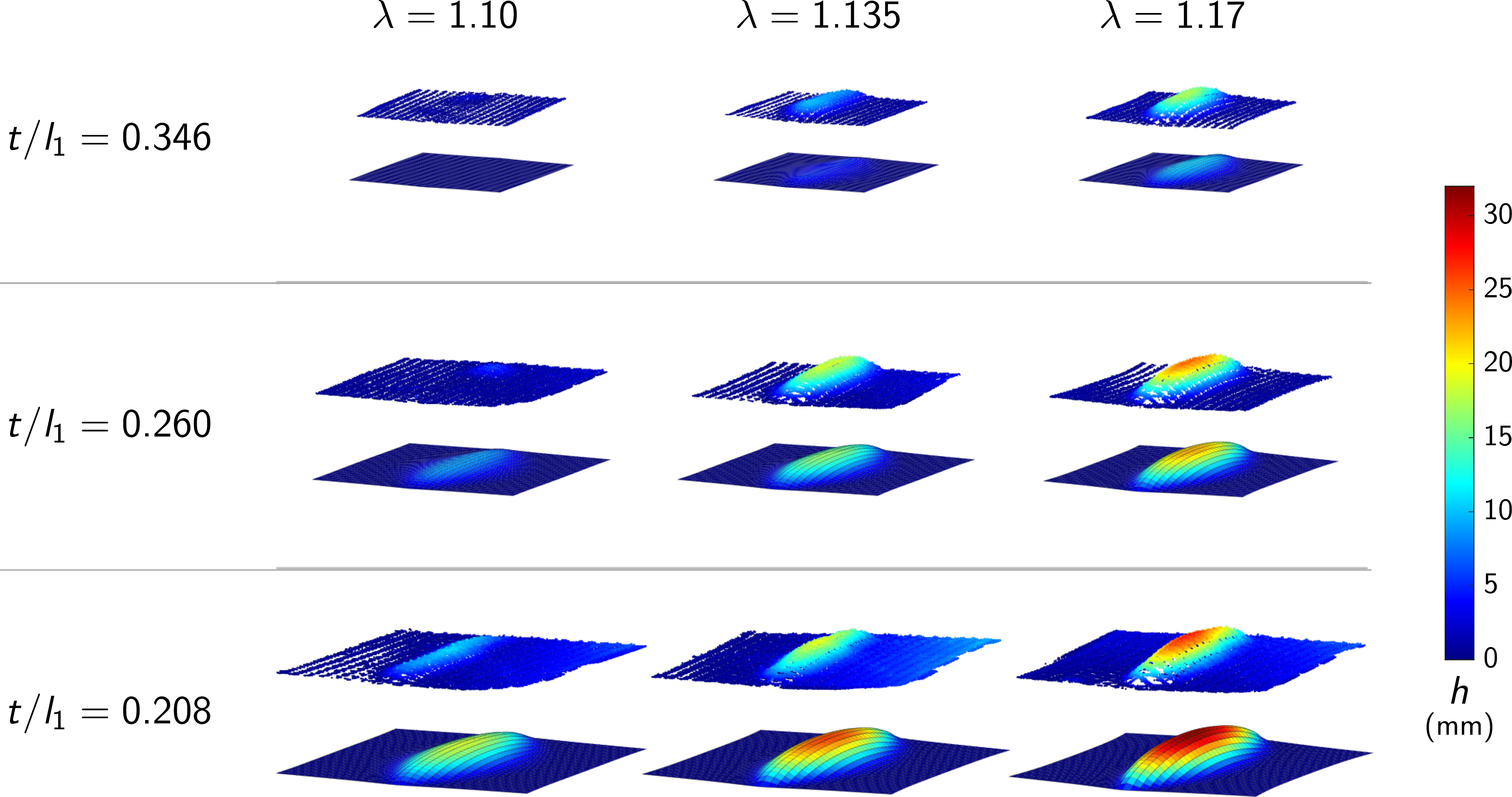}
\caption{Post-buckling behavior of sheets with three thickness-to-width ratios. These are the same three sheets represented in Fig.~\ref{f:heights}c. Here, $t$ is the sheet thickness, $l_1$ is the length of the unit cell grid spacing in the $\mathbf{e}_1$ direction, and $\lambda$ is the applied stretch at the midpoint of the sheet edge. In each entry of the stretch vs. aspect ratio grid, the laser scans are plotted directly above the simulated deformed meshes. As expected, we see that sheets with higher thickness-to-width ratios will nucleate at larger stretches and will buckle into wider domes relative to the overall sheet width.}
\label{f:laser}
\end{figure}

Finally, to better visualize how the post-buckling behavior evolves and is affected by the aspect ratio of the sheet, we show laser scans of the physical specimens and deformed simulation meshes at three different boundary point displacements in Fig.~\ref{f:laser}. Accurate quantitative comparisons are challenging due to the manual stitching process that follows the acquisition of laser scan data patches, which introduces slight distortions and puts certain regions of the scanned sheet at an inclined plane relative to the rest of the structure.  As expected, the post-buckled domes are wider (relative to the overall width of the sheets) for specimens that have larger thickness-to-width ratios, showing good qualitative agreement between experiments and simulations. Furthermore, the onset of buckling occurs at greater stretches as $t/l_1$ increases. 

These results show that this effective continuum modeling framework is a powerful tool for understanding the physics of quasi-mechanisms in non-periodic media. In our previous work~\cite{Celli2018}, we only captured in-plane deformation mappings using standard, fine-grained finite element procedures since the large number of elements needed to resolve the small mesostructural features (in the range between $10^5$ and $10^6$ elements depending on the structure being simulated) caused the calculation of out-of-plane buckling modes to have an inviable computational cost. Using the effective continuum approach we can get accurate results merely using a $36\times36$ uniform quadrilateral mesh, a reduction of two to three orders of magnitude in the number of elements used. Each of the bifurcation curves in Fig~\ref{f:heights} took roughly 5 minutes to compute running on a single core of a Intel\textsuperscript{\textregistered} Xeon\textsuperscript{\textregistered} 5218 processor. Meanwhile, we could not make simulations for the post-buckling behavior of our sheets converge in a reasonable amount of time using a standard fine-grained FEM approach.

\section{Conclusions}
\label{s:conclusions}

We present an effective continuum modeling framework for architected media that display quasi-mechanism behaviors and demonstrate its validity on sheets that are patterned with diamond-shaped cuts. The model incorporates a penalty for deviations from quasi-mechanism behaviors and relies on material model parameters extracted directly from experiments. We show that the approach correctly predicts the mechanical behavior of non-periodic media, even when the model's parameters are derived from experiments on periodic specimens. Our approach permits accurate and efficient simulations of mechanical behaviors that would otherwise be impractical to model using fine-grained simulations that fully resolve the material's small geometric features. 

We note that the implicit relation \eqref{eq:mechC} does not define the function $g(\mathbf{C},\boldsymbol{\phi})$ uniquely, implying that other choices of the functions $\Psi_p$ from \eqref{eq:energyPenQM} and $\Psi_m$ from \eqref{eq:memEnergy} are possible. A good agreement with experiments is still attained, suggesting that the buckling behavior of the sheet is robust with respect to the choice of the function $g$.

There are a few limitations to this approach. First, it requires a sufficient separation of length scales between the global deformation mode dimensions and the unit cell size. Therefore, it would not be able to capture the local buckling modes observed in some kirigami sheets~\cite{Rafsanjani2018a} or handle the dome kinking that occurs in our systems if they are fabricated from extremely thin sheets~\cite{Celli2018}. Furthermore, although we believe that this modeling approach could be applied to a broad range of architected media that display quasi-mechansims, extracting the material model constants from experiments may be more challenging in other systems in comparison to the perforated sheets we have discussed. Finding a suitable soft elastic energy density $\Psi_s$ that is appropriate for the quasi-mechanism regime also requires the modeler to have an intuition for which constitutive models can be appropriately tailored to fit experimental data attained from experiments on their system.

In the future, this modeling framework could be adapted to 3D media and materials with temporally varying mechanical properties, provided that they also display quasi-mechanisms.

\section*{Acknowledgements}

\noindent C.M. and C.D. were supported by the US Army Research Office Grant W911NF-17-1-0147. This work was also supported by a NASA Space Technology Research Fellowship to C.M. We thank Andrei Constantinescu and Kaushik Bhattacharya for helpful discussions, and Paul Stovall for assistance with fabrication. 

\bibliographystyle{elsarticle-num}
\bibliography{ShapeMBib}

\begin{thebibliography}{10}
\expandafter\ifx\csname url\endcsname\relax
  \def\url#1{\texttt{#1}}\fi
\expandafter\ifx\csname urlprefix\endcsname\relax\def\urlprefix{URL }\fi
\expandafter\ifx\csname href\endcsname\relax
  \def\href#1#2{#2} \def\path#1{#1}\fi

\bibitem{Schaedler2011}
T.~A. Schaedler, A.~J. Jacobsen, A.~Torrents, A.~E. Sorensen, J.~Lian, J.~R.
  Greer, L.~Valdevit, W.~B. Carter, Ultralight metallic microlattices, Science
  334~(6058) (2011) 962--965.

\bibitem{Ware2015}
T.~H. Ware, M.~E. McConney, J.~J. Wie, V.~P. Tondiglia, T.~J. White, Voxelated
  liquid crystal elastomers, Science 347~(6225) (2015) 982--984.
\newblock \href {https://doi.org/10.1126/science.1261019}
  {\path{doi:10.1126/science.1261019}}.

\bibitem{Moestopo2020}
W.~P. Moestopo, A.~J. Mateos, R.~M. Fuller, J.~R. Greer, C.~M. Portela, Pushing
  and pulling on ropes: Hierarchical woven materials, Advanced Science 7~(20)
  (2020) 2001271.
\newblock \href {https://doi.org/10.1002/advs.202001271}
  {\path{doi:10.1002/advs.202001271}}.

\bibitem{Rafsanjani2016}
A.~Rafsanjani, D.~Pasini, Bistable auxetic mechanical metamaterials inspired by
  ancient geometric motifs, Extreme Mechanics Letters 9 (2016) 291--296.
\newblock \href {https://doi.org/10.1016/j.eml.2016.09.001}
  {\path{doi:10.1016/j.eml.2016.09.001}}.

\bibitem{Klein2007}
Y.~Klein, E.~Efrati, E.~Sharon, {Shaping of Elastic Sheets by Prescription of
  Non-Euclidean Metrics}, Science 315~(5815) (2007) 1116--1120.
\newblock \href {https://doi.org/10.1126/science.1135994}
  {\path{doi:10.1126/science.1135994}}.

\bibitem{Gladman2016}
A.~S. Gladman, E.~A. Matsumoto, R.~G. Nuzzo, L.~Mahadevan, J.~A. Lewis,
  Biomimetic 4d printing, Nat. Mater. 15 (2016) 413--418.
\newblock \href {https://doi.org/10.1038/nmat4544}
  {\path{doi:10.1038/nmat4544}}.

\bibitem{Plucinsky2018}
P.~Plucinsky, B.~A. Kowalski, T.~J. White, K.~Bhattacharya, Patterning
  nonisometric origami in nematic elastomer sheets, Soft Matter 14 (2018)
  3127--3134.
\newblock \href {https://doi.org/10.1039/C8SM00103K}
  {\path{doi:10.1039/C8SM00103K}}.

\bibitem{Schenk2014}
M.~Schenk, A.~D. Viquerat, K.~A. Seffen, S.~D. Guest, Review of inflatable
  booms for deployable space structures: packing and rigidization, Journal of
  Spacecraft and Rockets 51~(3) (2014) 762--778.

\bibitem{Boley2019}
J.~W. Boley, W.~M. van Rees, C.~Lissandrello, M.~N. Horenstein, R.~L. Truby,
  A.~Kotikian, J.~A. Lewis, L.~Mahadevan, Shape-shifting structured lattices
  via multimaterial 4d printing, Proceedings of the National Academy of
  Sciences 116~(42) (2019) 20856--20862.
\newblock \href {https://doi.org/10.1073/pnas.1908806116}
  {\path{doi:10.1073/pnas.1908806116}}.

\bibitem{Greenberg2011}
H.~Greenberg, M.~Gong, S.~Magleby, L.~Howell, Identifying links between origami
  and compliant mechanisms, Mechanical Sciences 2~(2) (2011) 217--225.

\bibitem{Celli2020}
P.~Celli, A.~Lamaro, C.~McMahan, P.~Bordeenithikasem, D.~Hofmann, C.~Daraio,
  Compliant morphing structures from twisted bulk metallic glass ribbons,
  Journal of the Mechanics and Physics of Solids 145 (2020) 104129.

\bibitem{Ferraro2021}
S.~Ferraro, S.~Pellegrino, Topology and shape optimization of ultrathin
  composite self-deployable shell structures with cutouts, AIAA Journal (2021)
  1--14.

\bibitem{Guest1994}
S.~Guest, S.~Pellegrino, The folding of triangulated cylinders, part i:
  geometric considerations, Journal of Applied Mechanics 61 (1994) 773--777.

\bibitem{Dudte2016}
L.~H. Dudte, E.~Vouga, T.~Tachi, L.~Mahadevan, Programming curvature using
  origami tessellations, Nature Materials 15~(5) (2016) 583--588.
\newblock \href {https://doi.org/10.1038/NMAT4540}
  {\path{doi:10.1038/NMAT4540}}.

\bibitem{Wang2017}
F.~Wang, X.~Guo, J.~Xu, Y.~Zhang, C.~Q. Chen, Patterning curved
  three-dimensional structures with programmable kirigami designs, Journal of
  Applied Mechanics 84~(6) (2017) 061007.
\newblock \href {https://doi.org/10.1115/1.4036476}
  {\path{doi:10.1115/1.4036476}}.

\bibitem{Celli2018}
P.~Celli, C.~McMahan, B.~Ramirez, A.~Bauhofer, C.~Naify, D.~Hofmann, B.~Audoly,
  C.~Daraio, Shape-morphing architected sheets with non-periodic cut patterns,
  Soft Matter 14 (2018) 9744--9749.
\newblock \href {https://doi.org/10.1039/C8SM02082E}
  {\path{doi:10.1039/C8SM02082E}}.

\bibitem{Hawkes2010}
E.~Hawkes, B.~An, N.~M. Benbernou, H.~Tanaka, S.~Kim, E.~D. Demaine, D.~Rus,
  R.~J. Wood, Programmable matter by folding, Proc. Natl. Acad. Sci. U.S.A.
  107~(28) (2010) 12441--12445.
\newblock \href {https://doi.org/10.1073/pnas.0914069107}
  {\path{doi:10.1073/pnas.0914069107}}.

\bibitem{Shang2018}
X.~Shang, L.~Liu, A.~Rafsanjani, D.~Pasini, Durable bistable auxetics made of
  rigid solids, Journal of Materials Research 33~(3) (2018) 300--308.

\bibitem{Siefert2020}
E.~Si{\'e}fert, E.~Reyssat, J.~Bico, B.~Roman, Programming stiff inflatable
  shells from planar patterned fabrics, Soft Matter 16~(34) (2020) 7898--7903.

\bibitem{Guseinov2020}
R.~Guseinov, C.~McMahan, J.~P{\'e}rez, C.~Daraio, B.~Bickel, Programming
  temporal morphing of self-actuated shells, Nature Communications 11~(1)
  (2020) 1--7.
\newblock \href {https://doi.org/10.1038/s41467-019-14015-2}
  {\path{doi:10.1038/s41467-019-14015-2}}.

\bibitem{Agnelli2021}
F.~Agnelli, M.~Tricarico, A.~Constantinescu, Shape-shifting panel from 3d
  printed undulated ribbon lattice, Extreme Mechanics Letters 42 (2021) 101089.

\bibitem{Bertoldi2017}
K.~Bertoldi, V.~Vitelli, J.~Christensen, M.~Van~Hecke, Flexible mechanical
  metamaterials, Nature Reviews Materials 2~(11) (2017) 1--11.

\bibitem{Singh2021}
N.~Singh, M.~van Hecke, Design of pseudo-mechanisms and multistable units for
  mechanical metamaterials, Physical Review Letters 126 (2021) 248002.
\newblock \href {https://doi.org/10.1103/PhysRevLett.126.248002}
  {\path{doi:10.1103/PhysRevLett.126.248002}}.

\bibitem{Liu2019}
K.~Liu, T.~Tachi, G.~H. Paulino, Invariant and smooth limit of discrete
  geometry folded from bistable origami leading to multistable metasurfaces,
  Nature Communications 10~(1) (2019) 1--10.
\newblock \href {https://doi.org/10.1038/s41467-019-11935-x}
  {\path{doi:10.1038/s41467-019-11935-x}}.

\bibitem{Callens2017}
S.~J.~P. Callens, A.~A. Zadpoor, From flat sheets to curved geometries: Origami
  and kirigami approaches, Mater. Today 21~(3) (2018) 241--264.
\newblock \href {https://doi.org/10.1016/j.mattod.2017.10.004}
  {\path{doi:10.1016/j.mattod.2017.10.004}}.

\bibitem{Castle2014}
T.~Castle, Y.~Cho, X.~Gong, E.~Jung, D.~M. Sussman, S.~Yang, R.~D. Kamien,
  Making the cut: Lattice kirigami rules, Physical Review Letters 113 (2014)
  245502.
\newblock \href {https://doi.org/10.1103/PhysRevLett.113.245502}
  {\path{doi:10.1103/PhysRevLett.113.245502}}.

\bibitem{Tang2017}
Y.~Tang, J.~Yin, Design of cut unit geometry in hierarchical kirigami-based
  auxetic metamaterials for high stretchability and compressibility, Extreme
  Mechanics Letters 12 (2017) 77--85.
\newblock \href {https://doi.org/10.1016/j.eml.2016.07.005}
  {\path{doi:10.1016/j.eml.2016.07.005}}.

\bibitem{Jiang2020}
C.~Jiang, F.~Rist, H.~Pottmann, J.~Wallner, Freeform quad-based kirigami, ACM
  Transactions on Graphics 39~(6) (2020).
\newblock \href {https://doi.org/10.1145/3414685.3417844}
  {\path{doi:10.1145/3414685.3417844}}.

\bibitem{Grima2007}
J.~N. Grima, V.~Zammit, R.~Gatt, A.~Alderson, K.~E. Evans, Auxetic behaviour
  from rotating semi-rigid units, Physica Status Solidi B 244~(3) (2007)
  866--882.
\newblock \href {https://doi.org/10.1002/pssb.200572706}
  {\path{doi:10.1002/pssb.200572706}}.

\bibitem{Bertoldi2010}
K.~Bertoldi, P.~M. Reis, S.~Willshaw, T.~Mullin, Negative poisson's ratio
  behavior induced by an elastic instability, Advanced materials 22~(3) (2010)
  361--366.

\bibitem{Konakovic2016}
M.~Konakovi\'{c}, K.~Crane, B.~Deng, S.~Bouaziz, D.~Piker, M.~Pauly, Beyond
  developable: Computational design and fabrication with auxetic materials, ACM
  Transactions on Graphics 35~(4) (2016) 89.
\newblock \href {https://doi.org/10.1145/2897824.2925944}
  {\path{doi:10.1145/2897824.2925944}}.

\bibitem{Konakovic2018}
M.~Konakovi\'{c}-Lukovi\'{c}, J.~Panetta, K.~Crane, M.~Pauly, Rapid deployment
  of curved surfaces via programmable auxetics, ACM Transactions on Graphics
  37~(4) (2018) 106.
\newblock \href {https://doi.org/10.1145/3197517.3201373}
  {\path{doi:10.1145/3197517.3201373}}.

\bibitem{Overvelde2017}
J.~T. Overvelde, J.~C. Weaver, C.~Hoberman, K.~Bertoldi, Rational design of
  reconfigurable prismatic architected materials, Nature 541~(7637) (2017)
  347--352.

\bibitem{Gauss1828}
C.~Gauss, Disquisitiones generales circa superficies curvas, Typis
  Ditericianis, 1828.

\bibitem{Allaire2012}
G.~Allaire, A brief introduction to homogenization and miscellaneous
  applications, ESAIM: Proceedings 37 (2012) 1--49.
\newblock \href {https://doi.org/10.1051/proc/201237001}
  {\path{doi:10.1051/proc/201237001}}.

\bibitem{Muller1993}
S.~Müller, N.~Triantafyllidis, G.~Geymonat, Homogenization of nonlinearly
  elastic materials, microscopic bifurcation and macroscopic loss of rank-one
  convexity, Archive for Rational Mechanics and Analysis 122 (09 1993).
\newblock \href {https://doi.org/10.1007/BF00380256}
  {\path{doi:10.1007/BF00380256}}.

\bibitem{Schenk2011}
M.~Schenk, S.~D. Guest, Origami folding: A structural engineering approach,
  Origami 5 (2011) 291--304.

\bibitem{Filipov2017}
E.~T. Filipov, K.~Liu, T.~Tachi, M.~Schenk, G.~H. Paulino, Bar and hinge models
  for scalable analysis of origami, International Journal of Solids and
  Structures 124 (2017) 26--45.
\newblock \href {https://doi.org/10.1016/j.ijsolstr.2017.05.028}
  {\path{doi:10.1016/j.ijsolstr.2017.05.028}}.

\bibitem{Liu2017}
K.~Liu, G.~H. Paulino, Nonlinear mechanics of non-rigid origami: an efficient
  computational approach, Proceedings of the Royal Society A: Mathematical,
  Physical and Engineering Sciences 473~(2206) (2017) 20170348.
\newblock \href {https://doi.org/10.1098/rspa.2017.0348}
  {\path{doi:10.1098/rspa.2017.0348}}.

\bibitem{Hayakawa2020}
K.~Hayakawa, M.~Ohsaki, Form generation of rigid origami for approximation of a
  curved surface based on mechanical property of partially rigid frames,
  International Journal of Solids and Structures (2020).
\newblock \href {https://doi.org/10.1016/j.ijsolstr.2020.12.007}
  {\path{doi:10.1016/j.ijsolstr.2020.12.007}}.

\bibitem{Coulais2018}
C.~Coulais, C.~Kettenis, M.~{van Hecke}, A characteristic length scale causes
  anomalous size effects and boundary programmability in mechanical
  metamaterials, Nature Physics 14~(1) (2018) 40--44.
\newblock \href {https://doi.org/10.1038/nphys4269}
  {\path{doi:10.1038/nphys4269}}.

\bibitem{Leimer2020}
K.~Leimer, P.~Musialski, Reduced-order simulation of flexible meta-materials,
  in: Symposium on Computational Fabrication, SCF '20, Association for
  Computing Machinery, New York, NY, USA, 2020.
\newblock \href {https://doi.org/10.1145/3424630.3425411}
  {\path{doi:10.1145/3424630.3425411}}.

\bibitem{Baek2018}
C.~Baek, A.~O. Sageman-Furnas, M.~K. Jawed, P.~M. Reis, Form finding in elastic
  gridshells, Proceedings of the National Academy of Sciences 115~(1) (2018)
  75--80.
\newblock \href {https://doi.org/10.1073/pnas.1713841115}
  {\path{doi:10.1073/pnas.1713841115}}.

\bibitem{Lestringant2020}
C.~Lestringant, D.~M. Kochmann, Modeling of flexible beam networks and morphing
  structures by geometrically exact discrete beams, Journal of Applied
  Mechanics 87~(8) (2020) 081006.
\newblock \href {https://doi.org/10.1115/1.4046895}
  {\path{doi:10.1115/1.4046895}}.

\bibitem{yu2020numerical}
T.~Yu, L.~Dreier, F.~Marmo, S.~Gabriele, S.~Parascho, S.~Adriaenssens,
  Numerical modeling of static equilibria and bifurcations in elastic strip
  networks (2020).
\newblock \href {http://arxiv.org/abs/2011.06905} {\path{arXiv:2011.06905}}.

\bibitem{Reis2018}
P.~M. Reis, F.~Brau, P.~Damman, The mechanics of slender structures, Nature
  Physics 14~(12) (2018) 1150--1151.
\newblock \href {https://doi.org/10.1038/s41567-018-0369-4}
  {\path{doi:10.1038/s41567-018-0369-4}}.

\bibitem{BarSinai2020}
Y.~Bar-Sinai, G.~Librandi, K.~Bertoldi, M.~Moshe, Geometric charges and
  nonlinear elasticity of two-dimensional elastic metamaterials, Proceedings of
  the National Academy of Sciences 117~(19) (2020) 10195--10202.
\newblock \href {https://doi.org/10.1073/pnas.1920237117}
  {\path{doi:10.1073/pnas.1920237117}}.

\bibitem{Czajkowski2021}
M.~Czajkowski, C.~Coulais, M.~van Hecke, D.~Rocklin, Conformal elasticity of
  mechanism-based metamaterials, arXiv preprint arXiv:2103.12683 (2021).

\bibitem{Khajehtourian2021}
R.~Khajehtourian, D.~M. Kochmann, A continuum description of substrate-free
  dissipative reconfigurable metamaterials, Journal of the Mechanics and
  Physics of Solids 147 (2021) 104217.

\bibitem{Choi2019}
G.~P.~T. Choi, L.~H. Dudte, L.~Mahadevan, Programming shape using kirigami
  tessellations, Nature Materials 18~(9) (2019) 999--1004.
\newblock \href {https://doi.org/10.1038/s41563-019-0452-y}
  {\path{doi:10.1038/s41563-019-0452-y}}.

\bibitem{Jin2020}
L.~Jin, A.~E. Forte, B.~Deng, A.~Rafsanjani, K.~Bertoldi, Kirigami-inspired
  inflatables with programmable shapes, Advanced Materials 32~(33) (2020)
  2001863.
\newblock \href {https://doi.org/10.1002/adma.202001863}
  {\path{doi:10.1002/adma.202001863}}.

\bibitem{dealii}
D.~Arndt, W.~Bangerth, D.~Davydov, T.~Heister, L.~Heltai, M.~Kronbichler,
  M.~Maier, J.-P. Pelteret, B.~Turcksin, D.~Wells, The {deal.II} finite element
  library: Design, features, and insights, Computers \& Mathematics with
  Applications 81 (2021) 407--422.
\newblock \href {https://doi.org/10.1016/j.camwa.2020.02.022}
  {\path{doi:10.1016/j.camwa.2020.02.022}}.

\bibitem{Boffi2013}
D.~Boffi, F.~Brezzi, M.~Fortin, Mixed Finite Element Methods and Applications,
  Vol.~44, 2013.
\newblock \href {https://doi.org/10.1007/978-3-642-36519-5}
  {\path{doi:10.1007/978-3-642-36519-5}}.

\bibitem{Timoshenko1959}
S.~Timoshenko, S.~Woinowsky-Krieger, Theory of plates and shells (1959).

\bibitem{Shrimali2021}
B.~Shrimali, M.~Pezzulla, S.~Poincloux, P.~M. Reis, O.~Lopez-Pamies, The
  remarkable bending properties of perforated plates, Journal of the Mechanics
  and Physics of Solids (2021) 104514.

\bibitem{Rafsanjani2018a}
A.~Rafsanjani, Y.~Zhang, B.~Liu, S.~M. Rubinstein, K.~Bertoldi, Kirigami skins
  make a simple soft actuator crawl, Science Robotics 3~(15) (2018) eaar7555.
\newblock \href {https://doi.org/10.1126/scirobotics.aar7555}
  {\path{doi:10.1126/scirobotics.aar7555}}.

\end{thebibliography}

\newpage
\appendix
\section{Strong form of equilibrium} \label{app:strong}
The strong form of the equilibrium relations under the mixed formulation are
\begin{equation} \label{eq:strong_mixed}
	\begin{aligned}
		-\nabla \cdot \left( 2 \mathbf{F} \pdv{\Psi_m}{\mathbf{C}} \right) &= 0 \qquad && \text{in } \Omega, \\
		-\nabla \cdot \left( 2 \pdv{\Psi_m}{\mathbf{C}} \nabla w \right) + \pdv{\Psi_c}{w} + B \Delta v &= 0 && \text{in } \Omega, \\
		B(\Delta w - v) &= 0 \qquad && \text{in } \Omega, \\
	\end{aligned}
\end{equation}
with boundary conditions
\begin{equation}
	\begin{aligned}
		\left( 2 \mathbf{F} \pdv{\Psi_m}{\mathbf{C}} \right) \cdot n &= 0 \qquad && \text{on } \partial_f \Omega, \\
		u &= u_0 && \text{on } \partial_u \Omega, \\
		w = w_0, \ v &= 0 && \text{on } \partial \Omega.
	\end{aligned}
\end{equation}
The first two equations in \eqref{eq:strong_mixed} are the in-plane and out-of-plane momentum balance equations, respectively. The last equation is the constraint that $v = \Delta w$.

\section{Finite element formulation and Solution Procedure} \label{app:fem}
The fields $\mathbf{u}$, $w$, and $v$ are $H^1(\Omega)$, so we may consider a Galerkin finite element formulation with p = 1 shape functions for them. Therefore,
\begin{equation}
	\mathbf{u} = \sum_{i = 0}^{n_u} u_i \bm{\Phi}^u_i, \qquad w = \sum_{i = 0}^{n_w} w_i \Phi^w_i, \qquad  v = \sum_{i = 0}^{n_v} v_i \Phi^v_i,
\end{equation}
where $\{\bm{\Phi}^u_i\}$ is the set of vector-valued shape functions for the in-plane displacements. $\{\Phi^w_i\}$ and $\{\Phi^v_i\}$ are the scalar-valued sets of shape functions for $w$ and $v$, respectively. Because we assume homogeneous boundary conditions for both of these fields, we can then consider $\{\Phi^w_i\} = \{\Phi^v_i\}$. Then, using these shape functions for the variations in \eqref{eq:mixed_eq_weak}, the discrete equilibrium equations can be written as
\begin{equation}
	\begin{bmatrix}
		\mathbf{R}^u \\
		\mathbf{R}^w \\
		\mathbf{R}^v \\
	\end{bmatrix} = \mathbf{R} = \mathbf{0}\ ,
\end{equation}
where
\begin{equation}
	\begin{aligned}
		R^u_i &= \int_\Omega \left( 2 \mathbf{F} \pdv{\Psi_m}{\mathbf{C}} \right) : \nabla \bm{\Phi}^u_i \ dA, \\
		R^w_i &= \int_{\Omega} \left(  2 \pdv{\Psi_m}{\mathbf{C}} \nabla w  - B \nabla v \right) \cdot \nabla \Phi^w_i + \pdv{\Psi_c}{w} \Phi^w_i \ dA, \\
		R^v_i &= \int_{\Omega}  - B \, v \, \Phi^v_i - B \nabla w \cdot \nabla \Phi^v_i  \ dA. \\
	\end{aligned}
\end{equation}

To solve for this equilibrium, we use Newton-Raphson updates of the form
\begin{equation}
	\mathbf{K}(\mathbf{x}) \Delta \mathbf{x} = - \mathbf{R}(\mathbf{x}),
\end{equation}
where $\mathbf{x} = [ u_0, \ldots, u_{n_u}, w_0, \ldots, w_{n_w}, v_0, \ldots, v_{n_v} ]$ is the vector of degrees of freedom, $\Delta \mathbf{x}$ are their updates, and $\mathbf{K}$ is the tangent stiffness matrix
\begin{equation}
	\mathbf{K} = \begin{bmatrix}
		\mathbf{K}^{uu} & \mathbf{K}^{uw}  & \mathbf{0} \\
		\mathbf{K}^{wu} & \mathbf{K}^{ww}  & \mathbf{K}^{wv} \\
		\mathbf{0}      & \mathbf{K}^{vw}       & \mathbf{K}^{vv}
	\end{bmatrix},
\end{equation}
where
\begin{equation}
	\begin{aligned}
		K^{uu}_{ij} &= \int_\Omega \nabla \bm{\Phi}^u_i : \pdv{^2 \Psi_m}{\mathbf{F} \partial  \mathbf{F}} : \nabla \bm{\Phi}^u_j \ dA, \\
		K^{ww}_{ij} &= \int_\Omega \nabla \Phi^w_i \cdot \pdv{^2 \Psi_m}{\nabla w \partial \nabla w } \cdot \nabla \Phi^w_j \ dA, \\
		K^{vv}_{ij} &= \int_{\Omega} -B\, \Phi^v_i \Phi^v_j \ dA, \\
		K^{uw}_{ij} &= K^{wu}_{ji} = \int_\Omega \nabla \bm{\Phi}^u_i : \pdv{^2 \Psi_m}{\mathbf{F} \partial \nabla w } \cdot \nabla \Phi^w_j \ dA, \\
		K^{w v}_{ij} &= K^{v w}_{ji} = \int_\Omega -B \nabla \Phi^w_i \cdot \nabla \Phi^{v}_j \ dA. \\
	\end{aligned}
\end{equation} 
The displacements $\mathbf{u}_0$ on the boundary are incremented, and Newton-Raphson is used to reach an equilibrium configuration. The previous equilibrium configuration is used as an initial guess for the subsequent iterations.

\section{Stability analysis with mixed method constraint} \label{app:stab}
To probe the stability of an equilibrium configuration, it is common practice to calculate the eigenvalues of the tangent stiffness matrix. A negative eigenvalue implies an instability, and the equilibrium solution can be perturbed in the direction of the corresponding eigenvector to explore the buckled solution. In our case, we must restrict ourselves to eigenvectors in the subspace where the constraint $v = \Delta w$ is satisfied. To this end, we consider an effective stiffness matrix from the quadratic form, upon which the constraint is satisfied. Consider the discrete constraint equation:
\begin{equation}
	\mathbf{R}^{v} = \mathbf{K}^{v w} \mathbf{w} + \mathbf{K}^{v v} \mathbf{v} = \mathbf{0}.
\end{equation}
This can also be written in the following form:
\begin{equation}
	\mathbf{v} = -\left( \mathbf{K}^{v v} \right)^{-1} \mathbf{K}^{v w} \mathbf{w}.
\end{equation}
We can then use a reduced variable set $\mathbf{x}_r$ under which the constraint is satisfied, as
\begin{equation}
	\mathbf{x} = \begin{bmatrix}
		\mathbf{u} \\ 
		\mathbf{w} \\
		\mathbf{v} 
	\end{bmatrix} = \begin{bmatrix}
		\mathbf{I}_{n_u \times n_u} & \mathbf{0} \\ 
		\mathbf{0} & \mathbf{I}_{n_w \times n_w}\\
		\mathbf{0} & -\left( \mathbf{K}^{v v} \right)^{-1} \mathbf{K}^{v w}  
	\end{bmatrix} \begin{bmatrix}
		\mathbf{u} \\ 
		\mathbf{w}
	\end{bmatrix} = \mathbf{P} \mathbf{x}_r.
\end{equation}
Then, the quadratic form gives
\begin{equation}
	\mathbf{x}^T \mathbf{K} \mathbf{x} = \mathbf{x}_r^T \, \widetilde{\mathbf{K}} \, \mathbf{x}_r,
\end{equation}
where
\begin{equation}\label{eq:effStiffMat}
	\widetilde{\mathbf{K}} = \mathbf{P}^T \mathbf{K} \mathbf{P} = 
	\begin{bmatrix}
		\mathbf{K}^{uu} & \mathbf{K}^{uw} \\
		\mathbf{K}^{wu} & \left( \mathbf{K}^{ww} - \mathbf{K}^{w v} \left(  \mathbf{K}^{vv} \right)^{-1} \mathbf{K}^{v w} \right)
	\end{bmatrix}.
\end{equation}
Then to assess stability, we probe the eigenvalues of this effective stiffness matrix $\widetilde{\mathbf{K}}$. An eigenvalue passing through zero along the principle deformation path implies an instability. The corresponding eigenvector can then be used to produce a perturbation, using $\mathbf{P}$ to map back to the full variable set. The magnitude of the perturbation is chosen to be on the same order as the displacement increment. The direction of the perturbation is decided such that the $w$ component at the middle of the sheet is positive.

\end{document}